\documentclass[apsprd,showpacsx]{revtex4}
\usepackage{graphicx}
\usepackage{bm}
\usepackage{diagbox}
\usepackage{amsmath}
\usepackage{array}
\usepackage{color}
\usepackage{eufrak}
\usepackage{mathrsfs}
\usepackage{ulem}

\begin{document}

\title{ Meson Spectral Functions at Finite Temperature and Isospin Density with Functional Renormalization Group }
\author{Ziyue Wang and Pengfei Zhuang}
\affiliation{Physics Department, Tsinghua University and Collaborative Innovation Center of Quantum Matter, Beijing 100084, China}
\date{\today}
\begin{abstract}
The pion superfluid and the corresponding Goldstone and soft modes are investigated in two-flavor quark-meson model with functional renormalization group. By solving the flow equations for the effective potential and the meson two-point functions at finite temperature and isospin density, the critical temperature for the superfluid increases sizeably in comparison with solving the flow equation for the potential only. The spectral function for the soft mode shows clearly a transition from meson gas to quark gas with increasing temperature and a crossover from BEC to BCS pairing of quarks with increasing isospin density.
\end{abstract}
\maketitle

\section{Introduction}
\label{s1}
The Quantum Chromodynamics (QCD) phase transitions at finite temperature and density provide a deep insight into the strong interacting matter created in high energy nuclear collisions and compact stars, and the symmetry breaking and restoration patterns lead to a very rich QCD phase structure. The extension of the phase diagram from finite baryon density to isospin density is motivated by the investigation of isospin imbalance in the interior of neutron stars~\cite{Barshay:1974,Khodel:2004}. The thermodynamic equilibrium systems with finite isospin chemical potential have been widely investigated through perturbative QCD~\cite{Son:2000xc}, lattice simulations~\cite{Kogut:2002,Kogut:2004,deForcrand:2007,Detmold:2012,Endrodi:2014}, random matrix theory~\cite{Klein:2003, Klein:2004} and effective models like the Nambu--Jona-lasinio (NJL) model~\cite{He:2005}, quark-meson model~\cite{Kamikado:2012} and linear sigma model~\cite{Svanes:2010,Phat:2011}. It is found that the spontaneous breaking and restoration of the symmetry between charged pions $\pi_\pm$ are connected by a second order phase transition at both zero and finite temperature.

The BEC-BCS crossover, which is a hot topic in and beyond condensed matter and ultracold fermion gas, has been extended to relativistic fermion superfluid. Different from the normal situation where the BEC-BCS crossover is induced by increasing the attractive coupling among fermions~\cite{Eagles:1969zz,Nozieres:1985zz,SadeMelo:1993zz,Engelbrecht:1997zz,Babaev:1999iz,Castorina:2005tm,Diehl:2007ri,Nikolic:2007zz,Giorgini:2008zz}, there exists a BEC-BCS crossover in pion superfluid, controlled by the isospin density and chiral symmetry restoration~\cite{Sun:2007fc,He:2013gga}, similar to the case in two-color QCD system~\cite{Rezaeian:2006yj, Ferrer:2014ywa, Khan:2015puu}. In the superfluid phase close to the critical isospin density, the deconfinement does not yet happen, the system is in the BEC state of pions. On the other hand, at sufficiently high isospin density, the ground state of the system becomes a BCS superfluid. Between the two limits, the system has the same symmetry and is described by the same order parameter. Therefore, there should be a crossover from the BEC to BCS limits, instead of a phase transition, when the isospin chemical potential increases. It would naturally be intriguing to study the binding properties of the collective excitations in the system. Below the critical temperature, namely in the condensed phase, the global symmetry $O(2)$ is spontaneously broken with a Goldstone mode. Above the critical temperature, the Goldstone mode becomes a soft mode due to the thermal excitation. The study on the spectrum of the soft mode slightly above the phase transition line should give a way to understand the BEC-BCS crossover.

Considering the fact that mesons and quarks are dominant degrees of freedom in pion superfluid, we adopt in this paper the quark-meson model and focus on the spectrum of the soft mode in the BEC-BCS crossover. We employ the functional renormalization group (FRG)~\cite{Berges:2000ew,Polonyi:2001se,Pawlowski:2005xe,Gies:2006wv,Kopietz:2010zz, Braun:2011pp} approach to the quark-meson model. As a non-perturbative method, FRG enables us to incorporate fluctuation effects beyond mean field theory, see Refs.~\cite{Berges:2000ew}. The self-consistent treatment of fluctuations is important towards the understanding of physics near a phase transition. Since the FRG allows a description of scale transformation, it provides a deep insight into the system where scale dependence plays a crucial role. The FRG has been applied to a wide range of fields, including the QCD phase diagram at finite isospin chemical potential~\cite{Kamikado:2012,Svanes:2010} and in the two-color QCD~\cite{Kamikado:2012bt,Strodthoff:2013cua,Khan:2015puu}. The FRG has also been used to describe the BEC-BCS crossover in cold-atom systems, especially in the unitary limit~ \cite{Diehl:2007ri,Boettcher:2012cm,Boettcher:2012dh,Boettcher:2013kia}.

The spectral function encodes information about the degree of particle binding and the  collective excitation, and serves as the input for transport coefficients. To calculate the spectral function in the usually used imaginary time formalism with FRG, an analytical continuation is required to bring the imaginary time in the Euclidean two point function at finite temperature to the real time in the Minkowski space\cite{Pawlowski:2015mia,Strodthoff:2016pxx,Kamikado:2013sia,Tripolt:2013jra,Tripolt:2014wra,Jung:2016yxl}. This method has been applied to the study of real time observables such as shear viscosity \cite{Tripolt:2016cey} and soft modes\cite{Yokota:2016kyz} near the QCD critical point.

We organize the paper as follows. The FRG flow equations for the effective potential and meson two-poinr functions in the quark-meson model are derived in Section \ref{s2}. The procedure to solve the flow equations and the numerical results including the phase diagram and meson spectral functions are shown in Section \ref{s3}. We summarize in Section \ref{s4}.

\section{Flow Equations}
\label{s2}

As an low energy effective model, the quark-meson model comes from the partial bosonization of the four-fermion interaction model and exhibits many of the global symmetries of QCD. It is widely used as an effective chiral model to demonstrate the spontaneous chiral symmetry breaking in vacuum and its restoration at finite temperature and density~\cite{Jungnickel:1995fp,Schaefer:2004en, Pawlowski:2014zaa}. Here we take the two-flavor version of the model with pseudoscalar mesons ${\bf \pi}$ and scalar meson $\sigma$ as the dominant meson degrees of freedom at energy scale up to $\Lambda\approx 1$ GeV. The Euclidean effective action of the model at finite temperature $T$ is given as
\begin{equation}
\Gamma = \int_x\left[\bar\psi\left(i\partial\!\!\!/+m_0\right)\psi+ig\bar\psi\left(\sigma+i\gamma_5{\bf \tau}\cdot{\bf \pi}\right)\psi+\frac{1}{2}(\partial_\mu\phi)^2+U(\phi^2)-c\sigma\right],
\end{equation}
where the abbreviation $\int_x$ stands for $\int_0^\beta dx_0\int d^3x$ with the inverse of temperature $\beta=1/T$, and ${\bf \tau}$ are the Pauli matrices in flavor space. The Yukawa coupling is chosen as $g=3.2$ to fit the quark mass in vacuum, it is taken as a constant during the scale transformation. The fermion field $\psi$ and meson field $\phi$ are defined as $\psi=(u, d)$ and $\phi=(\sigma,\pi_+,\pi_-,\pi_0)$ in the complex representation. The real and complex representations are connected by a unitary transformation with $\pi_\pm=(\pi_1\pm i\pi_2)/\sqrt{2}$. The explicit chiral symmetry breaking term $-c\sigma$ corresponds to a finite current quark mass $m_0$.

In order to study the impact of finite isospin density, we introduce isospin chemical potential $\mu_I$ by adding a term $\mu_IQ_3$ with the associated conserved charge $Q_3$ to the corresponding field. Considering the isospin chemical potential $\pm \mu_I$ for quarks $u$ and $d$, the effective action of the system becomes
\begin{equation}
\Gamma = \int_x\left[\bar\psi S\psi+\left[(\partial_\mu+2\delta_{\mu 0}\mu_I)\pi_-\right]\left[(\partial_\mu-2\delta_{\mu 0}\mu_I)\pi_+\right]+\frac{1}{2}(\partial_\mu\sigma)^2+\frac{1}{2}(\partial_\mu\pi_0)^2+U(\phi^2)-c\sigma\right],
\end{equation}
where $S$ is the quark propagator in flavor space
\begin{equation}
\label{s}
S=\left(\begin{array}{cc}
M_+ & -\sqrt 2 g\gamma_5\pi_-\\
-\sqrt 2 g\gamma_5\pi_+ & M_-\\
\end{array}\right)
\end{equation}
with the diagonal elements $M_\pm =i(\partial\!\!\!/\pm\gamma_0\mu_I)+ig\left(\sigma\pm i\gamma_5\pi_0\right)$.

In vacuum at $\mu_I=0$, the $O(4)$ symmetry in chiral limit is explicitly broken to $O(3)$ with a rotational symmetry among the three pions in real case with $c\neq 0$. Turning on the isospin chemical potential leads to three distinguishable pions, and the $O(3)$ symmetry is explicitly broken to $O(2)$ symmetry. When $\mu_I$ exceeds the critical value $\mu_I^c=m_\pi/2$ with pion mass $m_\pi$ in vacuum, the symmetry is further spontaneously broken from $O(2)$ to $Z(2)$ and the system enters the pion superfluidity phase. With finite pion condensate, instead of the $O(4)$ invariant $\xi=\phi^2=\sigma^2+{\bf \pi}^2$, the effective potential $U(\rho,d)$ has separate dependence on two invariants $\rho=\sigma^2+\pi_0^2$ for the neutral mesons and $d=2\pi_+\pi_-=\pi_1^2+\pi_2^2$ for the charged pions.

Quantum and thermal fluctuations are of particular importance in the vicinity of a phase transition and are conveniently included within the framework of FRG. The core quantity in this approach is the averaged effective action $\Gamma_k$ at the RG scale $k$ in Euclidean space, its scale dependence is described by the flow equation~\cite{Berges:2000ew,Polonyi:2001se,Pawlowski:2005xe,Gies:2006wv,Kopietz:2010zz, Braun:2011pp}
\begin{equation}
\label{gamma}
\partial_k \Gamma_k =\text{Tr}\int_q\left[\frac{1}{2}G_{\phi,k}(p)R_{\phi,k}(p)-G_{\psi,k}(p)R_{\psi,k}(p)\right],
\end{equation}
where
\begin{eqnarray}
\label{g}
G_{\phi,k}(q) &=& \left[\Gamma_k^{(2)}[\phi]+R_{\phi,k}(q)\right]^{-1},\nonumber\\
G_{\psi,k}(q) &=& \left[\Gamma_k^{(2)}[\psi]+R_{\psi,k}(q)\right]^{-1}
\end{eqnarray}
are the FRG modified meson and quark propagators with the two-point functions $\Gamma_k^{(2)}[\phi]=\delta^2\Gamma_k/\delta\phi^2$ and $\Gamma_k^{(2)}[\psi]=\delta^2\Gamma_k/\delta\psi\delta\bar\psi$ and the two regulators $R_{\phi,k}$ and $R_{\psi,k}$. The symbol $\text{Tr}$ represents the summation over all inner degrees of freedom of mesons and quarks.  Assuming uniform field configurations, the integral over space and imaginary time becomes trivial, and the effective action $\Gamma_k=\beta V U_k$ is fully controlled by the potential $U_k$ with $V$ and $\beta$ being the space and time regions of the system.

The evolution of the flow from the ultraviolet limit $k=\Lambda$ to infrared limit $k=0$ encodes in principle all the quantum and thermal fluctuations in the action. To suppress the fluctuations with momentum smaller than the scale $k$ during the evolution, an infrared regulator $R$ is introduced in the flow equation. At finite temperature and density where the Lorentz symmetry is broken, we employ the optimized regulator function which is the three dimensional analogue of the 4-momentum regulator~\cite{Litim:2000,Litim:2001up,Litim:2002cf}. The bosonic and fermionic regulators are chosen to be
\begin{eqnarray}
R_{\phi,k}(p) &=& {\bf p}^2 r_B(y),\nonumber\\
R_{\psi,k}(p) &=& {\bf \gamma}\cdot {\bf p} r_F(y)
\end{eqnarray}
in momentum space with $y={\bf p}^2/k^2$ and $r_B(y)=(1/y-1)\Theta(1-y)$ and $r_F(y)=(1/\sqrt y-1)\Theta(1-y)$. The regulators $R_{\phi,k}$ and $R_{\psi,k}$ in the propagators $G_\phi$ and $G_\psi$ amount to having regularized three-momenta ${\bf p}_r^2={\bf p}^2(1+r_B(y))$ and ${\bf p}_r={\bf p}(1+r_F(y))$ for bosons and fermions respectively. The three dimensional regulators break down the Lorentz symmetry in vacuum. However, physical quantities are measured in the ground state at $k=0$, where the regulators vanish and the Lorentz symmetry is guaranteed.

We now derive the meson and quark propagators $G_\phi$ and $G_\psi$ in the flow equation for the effective potential $U_k$ in momentum space,
\begin{equation}
\label{uk}
\partial_k U_k = \text{Tr}\int_p\left[\frac{1}{2}G_{\phi,k}(p)R_{\phi,k}(p)-G_{\psi,k}(p)R_{\psi,k}(p)\right].
\end{equation}
We expand the effective potential around the mean field. Introducing the chiral condensate $\langle\sigma\rangle$ and pion condensate $\langle\pi\rangle$ to describe the chiral symmetry breaking and isospin symmetry breaking, and separating the meson field into a classical part $\phi_0=(\langle\sigma\rangle,\langle\pi\rangle/\sqrt 2, \langle\pi\rangle/\sqrt 2,0)$ and a quantum fluctuation part $\phi=(\sigma,\pi_+,\pi_-,\pi_0)$, the inverse of the FRG modified meson propagator in the superfluid phase with nonzero pion condensate is explicitly expressed as a matrix
\begin{equation}
G_{\phi,k}^{-1} = \left(\begin{array}{cccc}
H_\sigma & m_{\sigma\pi}^2 & m_{\sigma\pi}^2 & 0\\
m_{\sigma\pi}^2 & m_{\pi\pi}^2 & H_- & 0\\
m_{\sigma\pi}^2 &  H_+  & m_{\pi\pi}^2 & 0\\
0 & 0 & 0 &  H_0
\end{array}\right)
\end{equation}
with
\begin{eqnarray}
H_\sigma &=& {\bf p}_r^2+p_0^2+m_\sigma^2,\nonumber\\
H_0 &=& {\bf p}_r^2+p_0^2+m_{\pi_0}^2,\nonumber\\
H_\pm &=& {\bf p}_r^2+(p_0\mp 2i\mu_I)^2+m_\pi^2
\end{eqnarray}
and the curvature masses
\begin{eqnarray}
m_\sigma^2 &=& 2U_k^{(1,0)}+4\rho U_k^{(2,0)},\nonumber\\
m_{\pi_0}^2 &=& 2U_k^{(0,1)},\nonumber\\
m_\pi^2 &=& 2U_k^{(0,1)}+2d U_k^{(0,2)},\nonumber\\
m_{\sigma\pi}^2 &=& 2\sqrt{2\rho d}U^{(1,1)}_k,\nonumber\\
m_{\pi\pi}^2 &=& 2d U_k^{(0,2)}
\end{eqnarray}
determined by the derivatives of the potential with respect to the two invariants $U_k^{(m,n)}=\partial^m_\rho\partial^n_d U_k$. The propagator itself is then written as
\begin{equation}
\label{gphi}
G_{\phi,k} = \left(\begin{array}{cccc}
G_{\sigma} & G_{\sigma -} & G_{\sigma +} & 0\\
G_{\sigma +} & G_{+-} & G_+ & 0\\
G_{\sigma -} & G_-  & G_{+-} & 0\\
0 & 0 & 0 & G_0
\end{array}\right)
\end{equation}
with
\begin{eqnarray}
&& G_{\sigma} = (H_+H_--m_{\pi\pi}^4)/I,\\
&& G_0 = 1/H_0,\nonumber\\
&& G_\pm = (H_\mp H_\sigma-m_{\sigma\pi}^4)/I,\nonumber\\
&& G_{\sigma\pm} = m_{\sigma\pi}^2(m_{\pi\pi}^2-H_\mp)/I,\nonumber\\
&& G_{+-} = (m_{\sigma\pi}^4-m_{\pi\pi}^2H_\sigma)/I,\nonumber\\
&& I = H_\sigma(H_+H_--m_{\pi\pi}^4)-m_{\sigma\pi}^4(H_++H_--2m_{\pi\pi}^2).\nonumber
\end{eqnarray}

In the normal phase with vanishing pion condensate, the combination $\xi=\sigma^2+{\bf \pi}^2$ becomes the invariant of the system. In this case, the mixture among the meson fields disappears and the FRG modified meson propagator is simplified as
\begin{equation}
\label{meson_pro_n}
G_{\phi,k}=\left(\begin{array}{cccc}
G_{\sigma} & 0 & 0 & 0\\
0 & 0 & G_+ & 0\\
0 & G_-  & 0 & 0\\
0 & 0 & 0 & G_0
\end{array}\right)
\end{equation}
with
\begin{eqnarray}
G_{\sigma} &=& 1/H_{\sigma},\nonumber\\
G_\pm &=& 1/H_\pm,\nonumber\\
G_0 &=& 1/H_0.
\end{eqnarray}
Since all the off-diagonal elements in the meson mass matrix disappear in the normal phase, $m_{\sigma\pi}^2=m_{\pi\pi}^2=0$, the curvature masses are derived by the derivatives with respect to the invariant $\xi$,
\begin{eqnarray}
m_\sigma^2 &=& 2 U^{(1)}_k+4\xi U^{(2)}_k,\nonumber\\
m_{\pi_0}^2 &=& 2 U^{(1)}_k,\nonumber\\
m_{\pi_\pm}^2 &=& m_{\pi_0}^2\mp 2\mu_I.
\end{eqnarray}

The quark propagator is defined through the quark section in the effective action, see Eq.(\ref{s}). After a Fourier transformation at the expansion point, the quark propagator reads
\begin{equation}
S=\left(\begin{array}{cc}
p\!\!\!/+i\gamma_0\mu_I+ig\langle\sigma\rangle & -g\gamma_5\langle\pi\rangle\\
-g\gamma_5\langle\pi\rangle & p\!\!\!/-i\gamma_0\mu_I+ig\langle\sigma\rangle\\
\end{array}\right).
\end{equation}
From the definition (\ref{g}), the inverse of the FRG modified quark propagator is a matrix with off-diagonal elements in flavor space,
\begin{equation}
G_{\psi,k}^{-1}=\left(\begin{array}{cc}
\left(G_0^+\right)^{-1}& \Delta\\
\Delta & \left(G_0^-\right)^{-1}\\
\end{array}\right)
\end{equation}
with
\begin{eqnarray}
\left(G_0^\pm\right)^{-1} &=& {\bf p}\!\!\!/_r+\gamma_0(p_0\pm i\mu_I)+ig\langle\sigma\rangle\nonumber\\
\Delta &=& -\Delta_0\gamma_5, \quad \Delta_0=g\langle\pi\rangle.
\end{eqnarray}

In normal phase with $\langle\pi\rangle=0$ the modified propagator becomes diagonal,
\begin{eqnarray}
\label{quark_pro_n}
G_{\psi,k} &=& \left(\begin{array}{cc}
G_0^+& 0\\
0 & G_0^-\\
\end{array}\right),\nonumber\\
G_0^\pm &=& \frac{({\bf p}\!\!\!/-im_q)+(p_0+i\mu_I)\gamma_0}{(p_0\pm i\mu_I)^2+{\bf p}^2+m_q^2}
\end{eqnarray}
with quark mass $\left(m_q-m_0\right)^2=g^2\xi$. In general case with pion condensate we have
\begin{equation}
\label{gpsi}
G_{\psi,k}=\left(\begin{array}{cc}
G^+&\Xi^-\\
\Xi^+ & G^-\\
\end{array}\right)
\end{equation}
with diagonal and off-diagonal elements
\begin{eqnarray}
G^\pm &=& \left[(G_0^\pm)^{-1}-\Delta G_0^\mp\Delta\right]^{-1},\nonumber\\
\Xi^\pm &=& -G_0^\mp\Delta G^\pm.
\end{eqnarray}

With the three-dimensional regulator, the unregulated summation over Matsubara frequencies in (\ref{uk}) is performed analytically and the three-momentum integral is reduced to a trivial one. In normal phase, the flow equation is calculated at the scale-dependent expansion point $\phi_0=(\langle\sigma\rangle,0,0,0)$,
\begin{equation}
\label{normal}
\partial_k U_k = \frac{1}{2}\sum_\phi J_\phi(E_\phi,\mu_\phi)-N_c\sum_\psi J_\psi(E_\psi ,\mu_\psi ),
\end{equation}
where the meson and quark fields, chemical potentials and energies are defined as $\phi=\sigma, \pi_+, \pi_-, \pi_0, \psi =u, d,\ \mu_\sigma=0, \mu_+=2\mu_I, \mu_-=-2\mu_I, \mu_u=\mu_I, \mu_d=-\mu_I, E_\phi=\sqrt{{\bf p}^2+m_\phi^2}$ and $E_\psi =\sqrt{{\bf p}^2+m_\psi ^2}$ with meson and quark masses shown above, and the loop functions $J_\phi$ and $J_\psi$ are defined by
\begin{eqnarray}
J_{\phi} &=& \int_p \partial_k{R}_{\phi,k}(p)G_{\phi,k}(p),\nonumber\\
J_{\psi} &=& \int_p \partial_k R_{\psi,k}(p)G_{\psi,k}(p).
\end{eqnarray}
The momentum integral and Matsubara sum can be performed analytically, with the explicit expressions in the Appendix.

In the superfluid phase with both chiral and pion condensates, the flow equation is derived at the expansion point $\phi_0=(\langle\sigma\rangle,\langle\pi\rangle/\sqrt{2},\langle\pi\rangle/\sqrt{2},0)$,
\begin{equation}
\label{superfluid}
\partial_k U_k = \frac{1}{2}\sum_\phi J_\phi(E_\phi,0)-N_c\sum_\psi \left(1+{\mu_\psi \over E_\psi }\right) J_\psi(E_\psi ,0).
\end{equation}
Note that, when the isospin symmetry is spontaneously broken in the pion superfluid phase the normal mesons with definite isospin quantum numbers are no longer the eigenstates of the Hamiltonian of the system~\cite{He:2005}. The new eigenmodes are linear combination of these normal mesons. In two flavor case, the mixture is reflected in the off-diagonal elements of the meson and quark propagators (\ref{gphi}) and (\ref{gpsi}). While $\pi_0$ is not mixed with the other three mesons and still an eigenmode of the superfluid, $\sigma, \pi_+$ and $\pi_-$ are mixed with each other and the new eigenmodes $\Sigma, \Pi_+$ and $\Pi_-$ are their linear combinations. The energies $E_\phi$ of the quasi particles $\phi=\Sigma, \Pi_+, \Pi_-, \pi_0$ are defined at the four poles of the propagator $G_\phi$. By diagonalizing the quark propagator (\ref{gpsi}) the quark energies can be analytically written as $E_\psi=\sqrt{g^2\langle\pi\rangle^2+(\sqrt{{\bf p}^2+m_q^2}\pm\mu_I)^2}$ for $\psi=u, d$.

We now consider the flow equations for meson two-point functions. The purpose is to obtain the dressed propagators and the meson spectral functions, by including the decay channels of mesons and quarks. From the flow equation for the effective potential (\ref{uk}), we get only the statical properties of the mesons, namely the curvature masses defined through the effective potential. However, the meson masses in medium come not only from the background fields $\langle\sigma\rangle$ and $\langle\pi\rangle$ but also the interactions among mesons and quarks. From the two-point functions, one can extract the spectral functions which contain the information on full pole masses and decay properties of the mesons. At finite temperature and chemical potential, a propagator in medium depends separately on $p_0$ and ${\bf p}$, due to the breaking of Lorentz invariance. This leads to the difference between the pole mass and screening mass, defined through the pole of the propagator at vanishing three-momentum ${\bf p}=0$ and vanishing energy $p_0=0$, respectively. In the following we only consider the flow equations for two-point functions $\Gamma_{k,p}^{(2)}[\phi]$ at ${\bf p}=0$.

The flow equation for the two-point function $\Gamma^{(2)}_{k,p}[\phi_i]$ for meson $\phi_i$ with momentum $p$ is derived from the flow equation for the effective action $\Gamma_k[\phi,\psi]$ by taking its second order functional derivative with respect to $\phi_i$,
\begin{eqnarray}
\label{two-point}
\partial_k\Gamma^{(2)}_{k,p}[\phi_i] &=& \widetilde{\partial}_k\text{Tr}\int_q \Big[{1\over 2}G_{\phi,k}(q)\Gamma_k^{(4)}[\phi,\phi_i]-\frac{1}{2}G_{\phi,k}(q)\Gamma_k^{(3)}[\phi,\phi_i]G_{\phi,k}(q+p)\Gamma_k^{(3)}[\phi,\phi_i]\nonumber\\
&&\ \ \ \ \ \ \ \ \ \ \ \ +G_{\psi,k}(q)\Gamma_k^{(3)}[\psi,\phi_i]G_{\psi,k}(q+p)\Gamma_k^{(3)}[\psi,\phi_i]\Big],
\end{eqnarray}
where the symbol $\widetilde\partial_k$ means the derivative only on the regulators $R_k$.

Since we are interested in the thermal excitations above but close to the critical temperature of the pion superfluid, we investigate the meson two-point functions only in the normal phase, where the meson fields $\phi=(\sigma,{\bf \pi})$ are the eigenstates of the system and the propagators $G_\phi$ and $G_\psi$ in (\ref{two-point}) are presented in (\ref{meson_pro_n}) for mesons and (\ref{quark_pro_n}) for quarks.

In general, a flow equation for $i$-point function $\Gamma^{(i)}$ is coupled to the higher order functions $\Gamma^{(j)}$ with $j>i$, see the flow equations (\ref{gamma}) for $\Gamma$ which is related to $\Gamma^{(2)}$ and (\ref{two-point}) for $\Gamma^{(2)}$ which is related to $\Gamma^{(3)}$ and $\Gamma^{(4)}$. This leads to an infinite hierarchy of flow equations. The usually used way to truncate the hierarchy is to treat the higher order vertices as scale-dependent but momentum-independent couplings. That is the reason why we neglected the momentum dependence of $\Gamma^{(2)}$ in (\ref{gamma}) and of $\Gamma^{(3)}$ and $\Gamma^{(4)}$ in (\ref{two-point}). Under this truncation, the three-point and four-point functions or the three-line and four-line vertices are extracted from the global minimum of the effective potential. From the definition $\Gamma^{(3)}[\phi,\phi_i]=\delta\Gamma^{(2)}[\phi]/\delta\phi_i$, $\Gamma^{(3)}[\psi,\phi_i]=\delta \Gamma^{(2)}[\psi]/\delta\phi_i$ and $\Gamma^{(4)}[\phi,\phi_i]=\delta^2 \Gamma^{(2)}[\phi]/\delta\phi_i^2$ with external field $\phi_i$, we have in the normal phase the coupling matrices
\begin{eqnarray}
\Gamma^{(3)}_k[\phi,\sigma] &=& \left(\begin{array}{cccc}
\gamma^{(3,1)}_k& 0 & 0 & 0\\
0 & 0 & \gamma^{(3,2)}_k &0 \\
0 &  \gamma^{(3,2)}_k & 0 &0 \\
0 & 0 & 0 &\gamma^{(3,2)}_k
\end{array}\right),\nonumber
\qquad\qquad \Gamma^{(3)}_k[\phi,\pi_+] = \left(\begin{array}{cccc}
0 & 0 & \gamma^{(3,2)}_k  & 0\\
0 & 0 & 0 & 0\\
\gamma^{(3,2)}_k &0  & 0 & 0\\
0 &0  & 0 &0
\end{array}\right),\nonumber\\
~~~\\
\Gamma^{(3)}_k[\phi,\pi_-] &=& \left(\begin{array}{cccc}
0 & \gamma^{(3,2)}_k & 0 & 0\\
\gamma^{(3,2)}_k &0 & 0 & 0\\
0 & 0 & 0 & 0\\
0 & 0 & 0 &0
\end{array}\right),\nonumber
\qquad\qquad\qquad\qquad\Gamma^{(3)}_k[\phi,\pi_0] = \left(\begin{array}{cccc}
0 & 0 & 0 &  \gamma^{(3,2)}_k\\
0 & 0 & 0 & 0\\
0 & 0 & 0 & 0\\
\gamma^{(3,2)}_k &0  & 0 &0
\end{array}\right)
\end{eqnarray}
for the three-meson vertices with $\gamma_k^{(3,1)}=12\langle\sigma\rangle_k U_k^{(2)}+8\langle\sigma\rangle_k^3U^{(3)}_k$ and $\gamma_k^{(3,2)}=4\langle\sigma\rangle_k U^{(2)}_k$,
\begin{eqnarray}
\Gamma^{(4)}_k[\phi,\sigma] &=& \left(\begin{array}{cccc}
\gamma^{(4,1)}_k & 0 & 0 & 0\\
0 & 0 & \gamma^{(4,2)}_k &0 \\
0 &  \gamma^{(4,2)}_k & 0 &0 \\
0 & 0 & 0 &\gamma^{(4,2)}_k
\end{array}\right),\nonumber\\
\Gamma^{(4)}_k[\phi,\pi_\pm] &=& \left(\begin{array}{cccc}
\gamma^{(4,2)}_k & 0 & 0 & 0\\
0 & 0 & \gamma^{(4,3)}_k &0 \\
0 & \gamma^{(4,3)}_k & 0 &0 \\
0 & 0 & 0 &\gamma^{(4,4)}_k
\end{array}\right),\nonumber\\
\Gamma^{(4)}_k[\phi,\pi_0] &=& \left(\begin{array}{cccc}
\gamma^{(4,2)}_k & 0 & 0 & 0\\
0 & 0 & \gamma^{(4,4)}_k &0 \\
0 & \gamma^{(4,4)}_k & 0 &0 \\
0 & 0 & 0 &\gamma^{(4,5)}_k
\end{array}\right)
\end{eqnarray}
for the four-meson vertices with $\gamma_k^{(4,1)}=12U_k^{(2)}+48\langle\sigma\rangle_k^2U_k^{(3)}+16\langle\sigma\rangle_k^4U_k^{(4)}$, $\gamma_k^{(4,2)}=4U_k^{(2)}+8\langle\sigma\rangle_k^2U_k^{(3)}$, $\gamma_k^{(4,3)}=8U_k^{(2)}$, $\gamma_k^{(4,4)}=4U_k^{(2)}$ and $\gamma_k^{(4,5)}=12U_k^{(2)}$, and
\begin{eqnarray}
\Gamma^{(3)}_k[\psi,\sigma] &=& ig,\nonumber\\
\Gamma^{(3)}_k[\psi,\pi_0] &=& -g\gamma_5\left(\begin{array}{cc}
1 &  0\\
0 & -1
\end{array}\right),\nonumber\\
\Gamma^{(3)}_k[\psi,\pi_+] &=& -\sqrt{2}g\gamma_5\left(\begin{array}{cc}
0 & 0\\
1 & 0
\end{array}\right),\nonumber\\
\Gamma^{(3)}_k[\psi,\pi_-] &=& -\sqrt{2}g\gamma_5\left(\begin{array}{cc}
0 & 1\\
0 & 0
\end{array}\right)
\end{eqnarray}
for the meson-quark vertices.

\begin{figure}[!hbt]\centering
\includegraphics[height=0.18\textwidth]{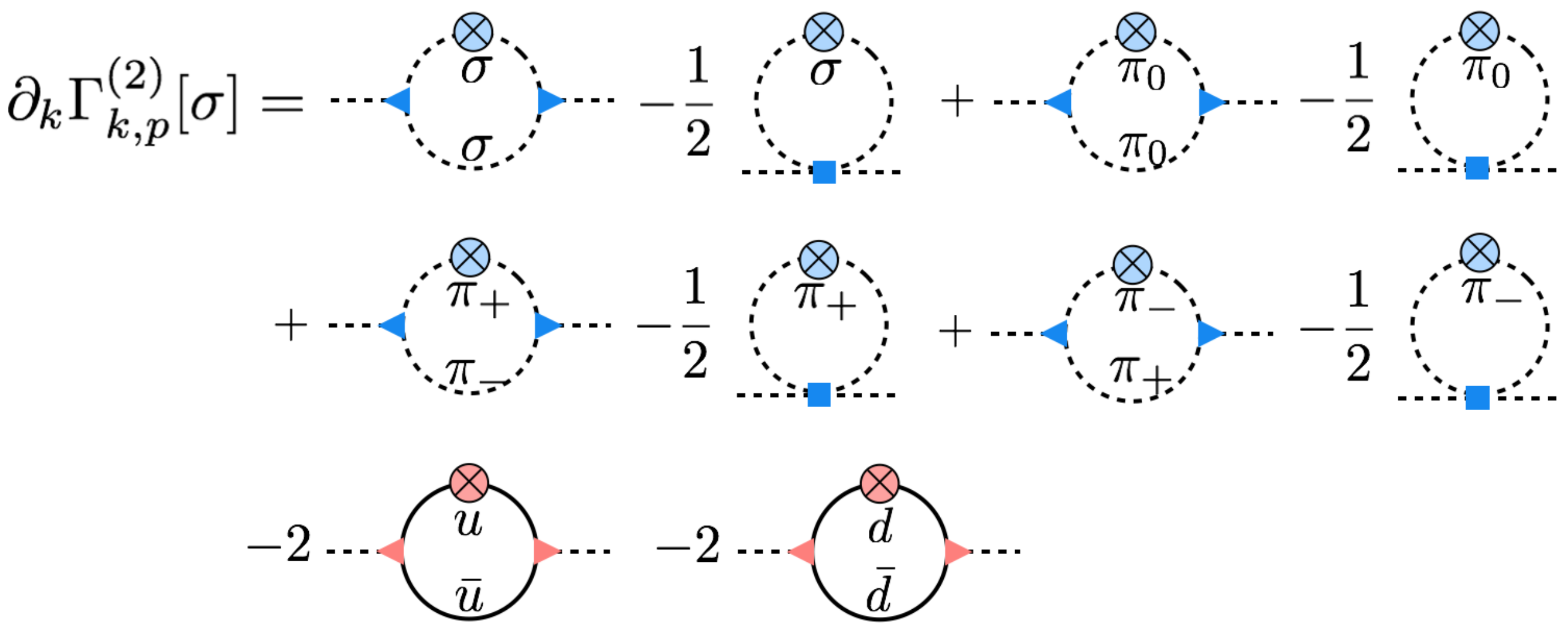}~~~~~~
\includegraphics[height=0.18\textwidth]{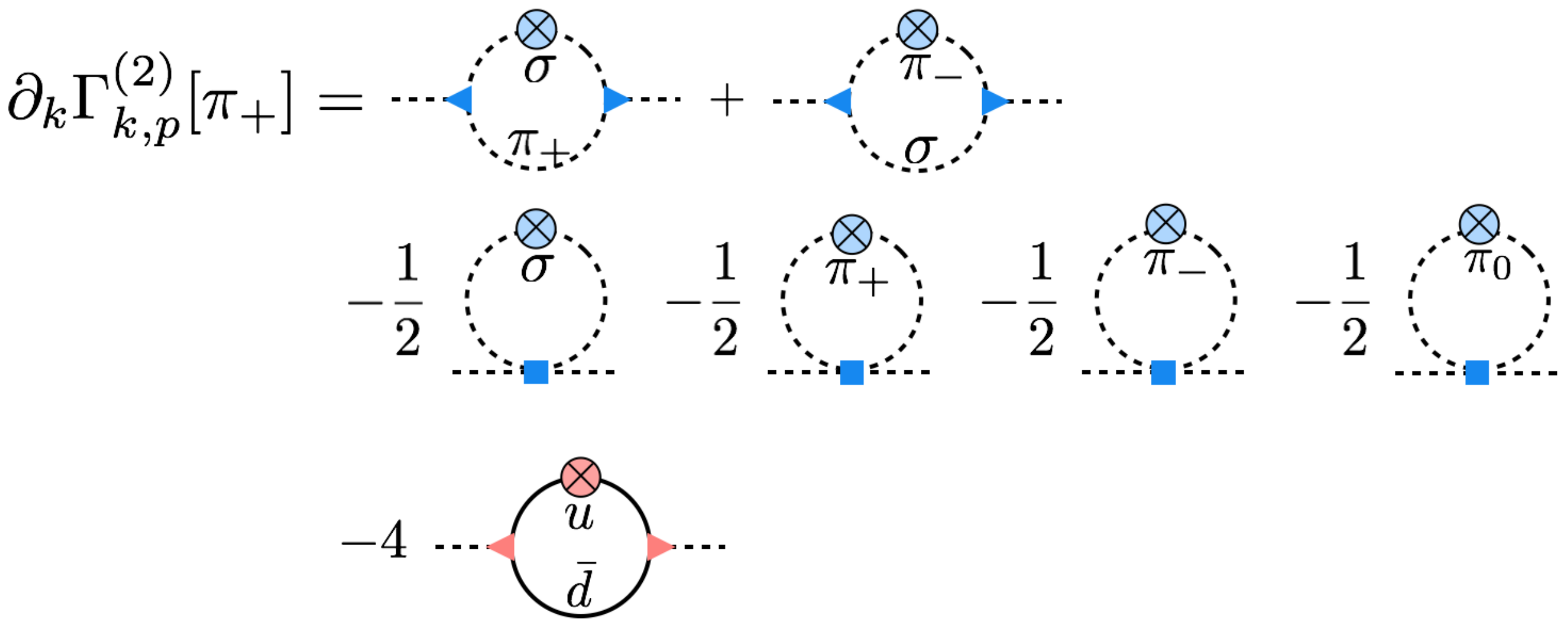}\\
~~\\
\includegraphics[height=0.18\textwidth]{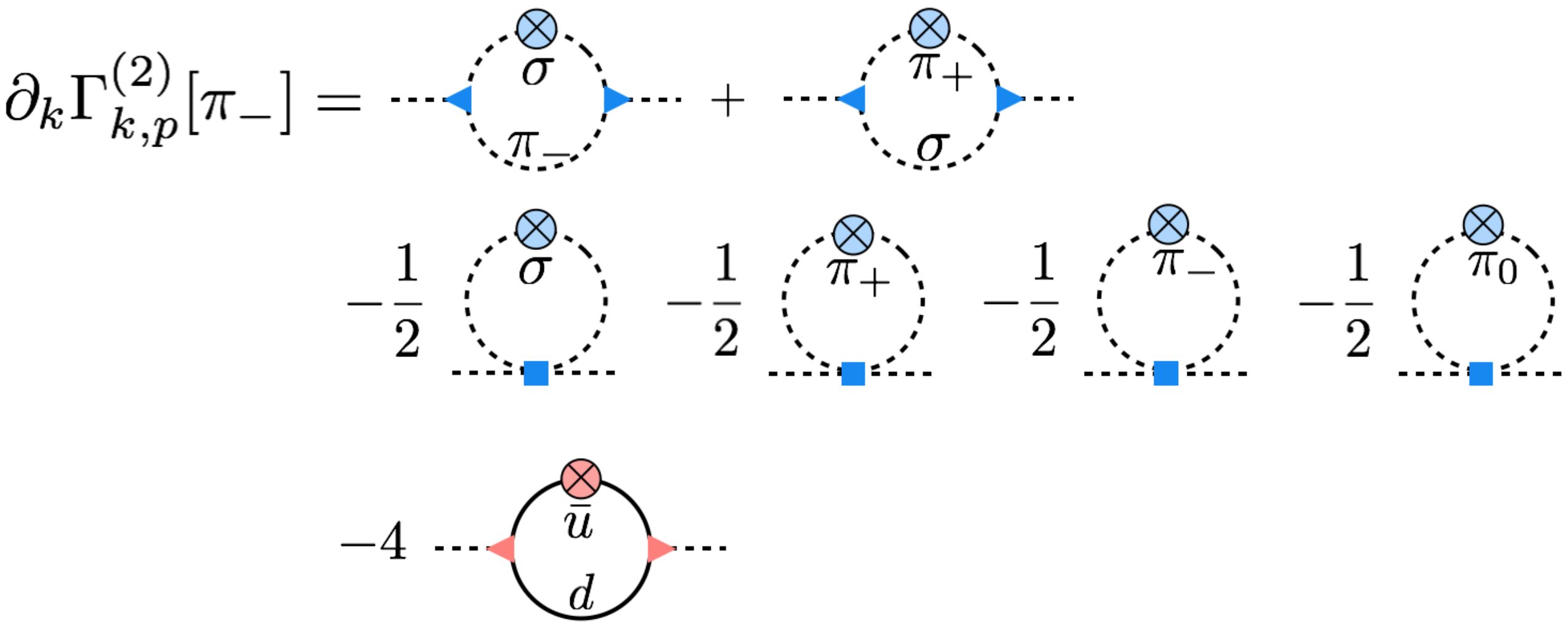}~~~~~~
\includegraphics[height=0.18\textwidth]{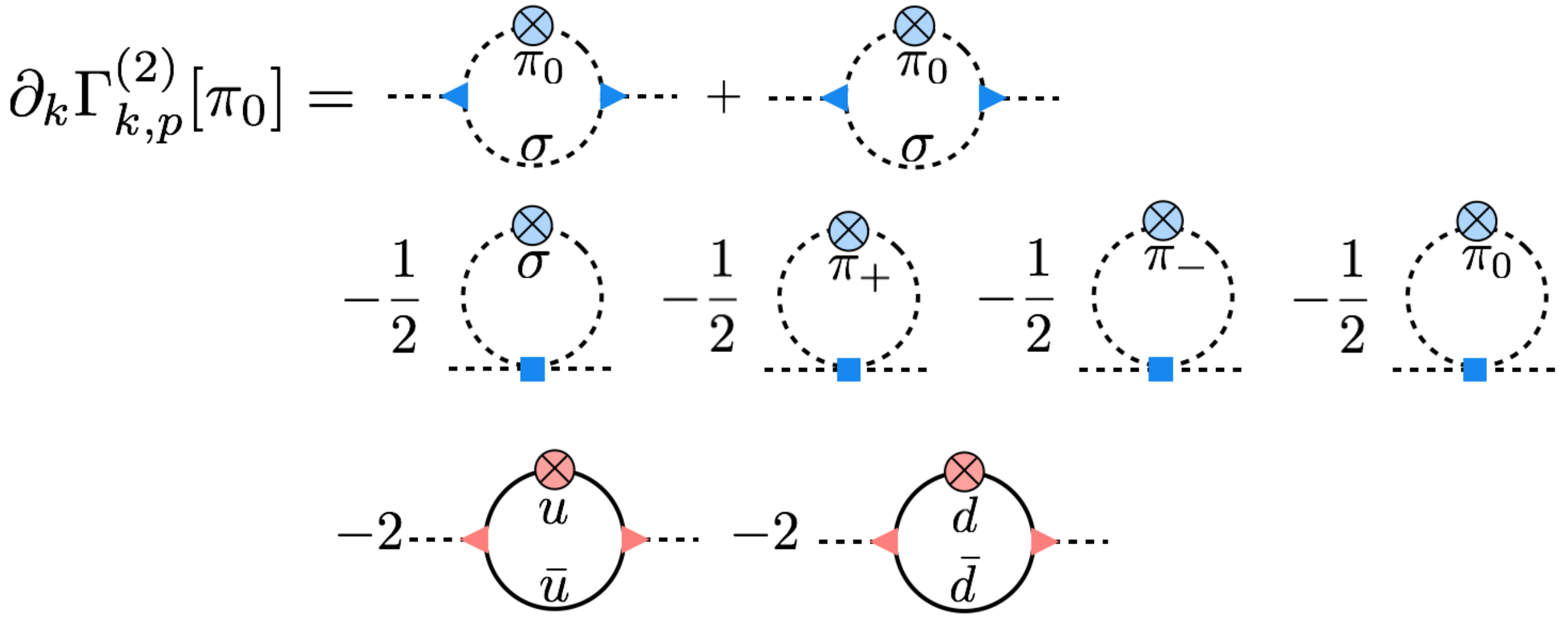}
\caption{ The diagrammatic representation of the flow equations for the meson two-point functions $\Gamma_{k,p}^{(2)}[\phi_i]$. The dashed and solid lines are the meson and quark fields, the triangles and squares indicate the three-line and four-line vertices, and $p$ is the momentum of the external field $\phi_i$.}
\label{fig1}
\end{figure}
With the known vertices, the flow equations for the two-point functions $\Gamma_{k,p}^{(2)}[\phi_i]$ can be diagrammatically shown in Fig.\ref{fig1}, where the three-line and four-line vertices are respectively represented by triangles and squares and $p$ is the momentum of the external field $\phi_i$.

In order to investigate the meson pole masses beyond the potential level, we evaluate the flow equations for the two-point functions at vanishing external three-momentum ${\bf p}=0$. In this case, the momentum integral and Matsubara sum involved in the one-loop Feynman diagrams can be performed analytically, giving the following explicit expression
\begin{eqnarray}
\partial_k\Gamma^{(2)}_{k,p_0}[\sigma]
&=& -\frac{1}{2}\left[\gamma_k^{(4,1)}K_\sigma+\gamma_k^{(4,2)}\big(K_0+K_++K_-\big)\right]
+ \left[\gamma_k^{(3,1)}\right]^2L_{\sigma\sigma}+\left[\gamma_k^{(3,2)}\right]^2\big(L_{00}+L_{++}+L_{--}\big)\nonumber\\
&&-6g^2\left(M^\sigma_{++}+M^\sigma_{--}\right),\nonumber\\
\partial_k\Gamma^{(2)}_{k,p_0}[\pi_+]
&=& -\frac{1}{2}\left[\gamma_k^{(4,2)}K_\sigma+\gamma_k^{(4,3)}\big(K_++K_-\big)+\gamma_k^{(4,4)}K_0\right]
+ \left[\gamma_k^{(3,2)}\right]^2\big(L_{\sigma -}+L_{+\sigma}\big)-12g^2M^\pi_{-+},\nonumber\\
 \partial_k\Gamma^{(2)}_{k,p_0}[\pi_-]
&=& -\frac{1}{2}\left[\gamma_k^{(4,2)}K_\sigma+\gamma_k^{(4,3)}\big(K_++K_-\big)+\gamma_k^{(4,4)}K_0\right]
+ \left[\gamma_k^{(3,2)}\right]^2\big(L_{-\sigma}+L_{\sigma +}\big)-12g^2M^\pi_{+-},\nonumber\\
 \partial_k\Gamma^{(2)}_{k,p_0}[\pi_0]
 &=& -\frac{1}{2}\left[\gamma_k^{(4,2)}K_\sigma+\gamma_k^{(4,4)}\big(K_++K_-\big)+\gamma_k^{(4,5)}K_0\right]
+\left[\gamma_k^{(3,2)}\right]^2\big(L_{\sigma 0}+L_{0\sigma}\big)-6g^2\big(M^\pi_{++}+M^\pi_{--}\big).
\label{twopoint}
\end{eqnarray}
The threshold functions $K_{\phi_i}$, $L_{\phi_i\phi_j}(p_0)$ and $M^{\phi_i}_{\phi_j\phi_k}(p_0)$ represent the momentum integral and Matsubara sum in the loops. $K_{\phi_i}$ is for the meson loops with four-line vertices,
\begin{equation}
K_{\phi_i}=\int_q \partial_k R_{\phi,k}(q)G^2_{\phi_i,k}(q),
\end{equation}
$L_{\phi_i\phi_j}$ is for the meson loops with three-line vertices,
\begin{equation}
L_{\phi_i\phi_j}(p_0)=\int_q \partial_kR_{\phi,k}(q)G^2_{\phi_i,k}(q)G_{\phi_j,k}(q+p),
\end{equation}
and $M^{\phi_i}_{\phi_j\phi_k}$ is related to the fermion loops,
\begin{equation}
 M^{\phi_i}_{\phi_j\phi_k}(p_0) = \text{Tr}\int_q\partial_kR_{\psi,k}(q)G_{\psi,k}(q)\Gamma^{(3)}_k[\psi,\phi_j]G_{\psi,k}(p+q)\Gamma^{(3)}_k[\psi,\phi_k]G_{\psi,k}(q),
\end{equation}
where the trace is done in flavor space and Dirac space. The momentum integral and Matsubara sum for the meson and quark loops can be done analytically. The explicit expressions are presented in the Appendix.

In order to perform the analytical continuation to obtain the two-point functions in Minkowski space, we make the analytic continuation
\begin{equation}
\Gamma^{(2)}_{k,\omega}[\phi_i] = \lim_{\epsilon\to 0}\lim_{p_0\to -i(\omega+i\epsilon)}\Gamma^{(2)}_{k,p_0}[\phi_i].
\end{equation}
This substitution of the discrete Euclidean frequency $p_0$ by the continuous energy $\omega$ is done explicitly before the integration of the RG scale $k$.

Finally, the meson spectral functions are expressed in terms of the imaginary and real parts of the retarded propagator,
\begin{equation}
\rho_{k,\omega}[\phi_i]=-\frac{1}{\pi}\frac{\text{Im}\Gamma^{(2)}_{k,\omega}[\phi_i]}{\left[\text{Re}\Gamma^{(2)}_{k,\omega}[\phi_i]\right]^2+\left[\text{Im}\Gamma^{(2)}_{k,\omega}[\phi_i]\right]^2}.
\end{equation}

\section{Numerical Treatment and Results}
\label{s3}

To numerically solve the flow equations for the effective potential and the two-point functions, we have to specify the model parameters and provide initial conditions. For the effective potential we assume the initial condition at the ultraviolet limit,
\begin{equation}
\label{ni}
U_\Lambda(\xi)=\frac{1}{2}m_\Lambda^2\xi+\frac{1}{4}\lambda_\Lambda\xi^2
\end{equation}
for one-dimensional grid in normal phase and
\begin{equation}
U_\Lambda(\rho,d)=\frac{1}{2}m_\Lambda^2(\rho+d)+\frac{1}{4}\lambda_\Lambda(\rho+d)^2
\end{equation}
for two-dimensional grid in pion superfluid. For solving the meson two-point functions in normal phase, we take (\ref{ni}) and
\begin{eqnarray}
\Gamma_{\Lambda,\omega}^{(2)}[\sigma] &=& -\omega^2+2U^{(1)}_\Lambda+4\phi^2U^{(2)}_\Lambda,\nonumber\\
\Gamma_{\Lambda,\omega}^{(2)}[\pi_\pm] &=& -(\omega\pm 2\mu_I)^2+2U^{(1)}_\Lambda,\nonumber\\
\Gamma_{\Lambda,\omega}^{(2)}[\pi_0] &=& -\omega^2+2U^{(1)}_\Lambda
\end{eqnarray}
as the initial condition. The parameters $m_\Lambda$ and $\lambda_\Lambda$ and the scale independent parameter $c$ are fixed by fitting the meson and quark masses in vacuum at the infrared limit $k=0$ of the flow equations. We calculate the meson masses in two cases, solving only the flow equation for the potential and solving the flow equations for both the potential and meson two-point functions. In the former case (case A), the pole masses are just the curvature masses determined by the flow equations (\ref{normal}) and (\ref{superfluid}) for the potential. In the latter case (case B), the pole masses are extracted from the corresponding spectral functions controlled by the flow equations (\ref{twopoint}) for the two-point functions, where the potential solved from (\ref{normal}) is used to determine the vertices in the two-point functions. During the process of integrating the flow equations from the ultraviolet limit to the infrared limit, the condensates $\langle\sigma\rangle_k$ and $\langle\pi\rangle_k$ are obtained by locating the minimum of the $k$-dependent effective potential $U_k$. Choosing the quark mass $m_q=300$ MeV, pion mass $m_\pi=134$ MeV and pion decay constant $f_\pi=94$ MeV in vacuum, the corresponding initial parameters are $m_\Lambda^2/\Lambda^2=0.618$, $\lambda_\Lambda=1$ and $c/\Lambda^3=0.0023$ in case A and $m_\Lambda^2/\Lambda^2=0.645$, $\lambda_\Lambda=1$ and $c/\Lambda^3=0.0045$ in case B with the cutoff $\Lambda=900$ MeV.

\begin{figure}[!hbt]\centering
\includegraphics[height=0.26\textwidth]{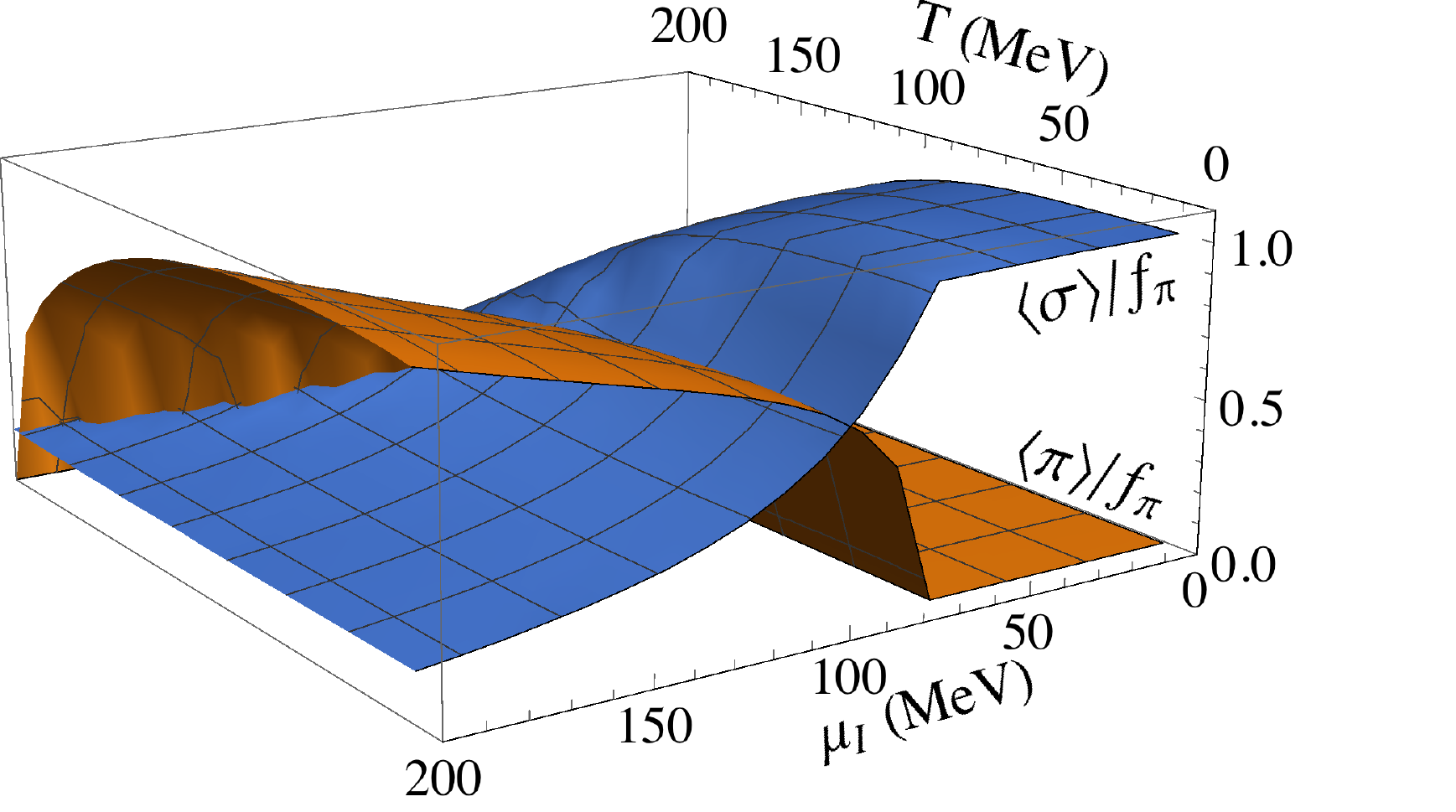}
\caption{The temperature and isospin chemical potential dependence of the chiral and pion condensates. }
\label{fig2}
\end{figure}
Considering the fact that the system in ultraviolet limit should be controlled by the dynamics and not sensitive to the external parameters like temperature $T$ and isospin chemical potential $\mu_I$, we take medium independent initial condition in solving the flow equations at finite $T$ and $\mu_I$~\cite{Berges:2000ew,Polonyi:2001se}. Fig.\ref{fig2} shows the $T$ and $\mu_I$ dependence of the chiral and pion condensates $\langle\sigma\rangle$ and $\langle\pi\rangle$, by integrating the flow equation for the potential from $\Lambda$ to $0$ with the initial condition A. Due to thermal excitation, both condensates are gradually melted in the hot medium and start to vanish at corresponding critical temperatures. The isospin density effect is totally different for the chiral and pion condensates. For the $u\bar u$ or $d\bar d$ pairing, the imbalance between the two Fermi surfaces increases linearly with $\mu_I$, but for the $u\bar d$ or $d\bar u$ pairing, the two Fermi surfaces are the same and the hight increases linearly with $\mu_I$. Therefore, the chiral condensate decreases but the pion condensate increases with $\mu_I$.

There are two ways to determine the phase boundary of the pion superfluid in the $T$ and $\mu_I$ plane, by approaching the boundary in the superfluid phase and in the normal phase. In the superfluid phase, the pion condensate is nonzero but approaches to zero when the system moves towards the phase boundary, namely the phase transition line is defined through the condition $\langle\pi\rangle(T,\mu_I)=0$. On the other hand, from the Goldstone theorem, there should be Goldstone modes when a global symmetry is spontaneously broken. From the discussions in Section \ref{s2}, $\Pi_+$ is the Goldstone mode in the superfluid phase, corresponding to the spontaneous isospin symmetry breaking. It is massless in the whole superfluid phase. Considering the continuity of the eigenmodes of the system on the phase boundary, $\pi_+$ in the normal phase should become massless on the boundary. Therefore, the phase transition line can also be defined through the condition $m_{\pi_+}(T,\mu_I)=0$. The above two definitions should be equivalent, guaranteed by the Goldstone theorem.

\begin{figure}[!hbt]\centering
\includegraphics[width=0.44\textwidth]{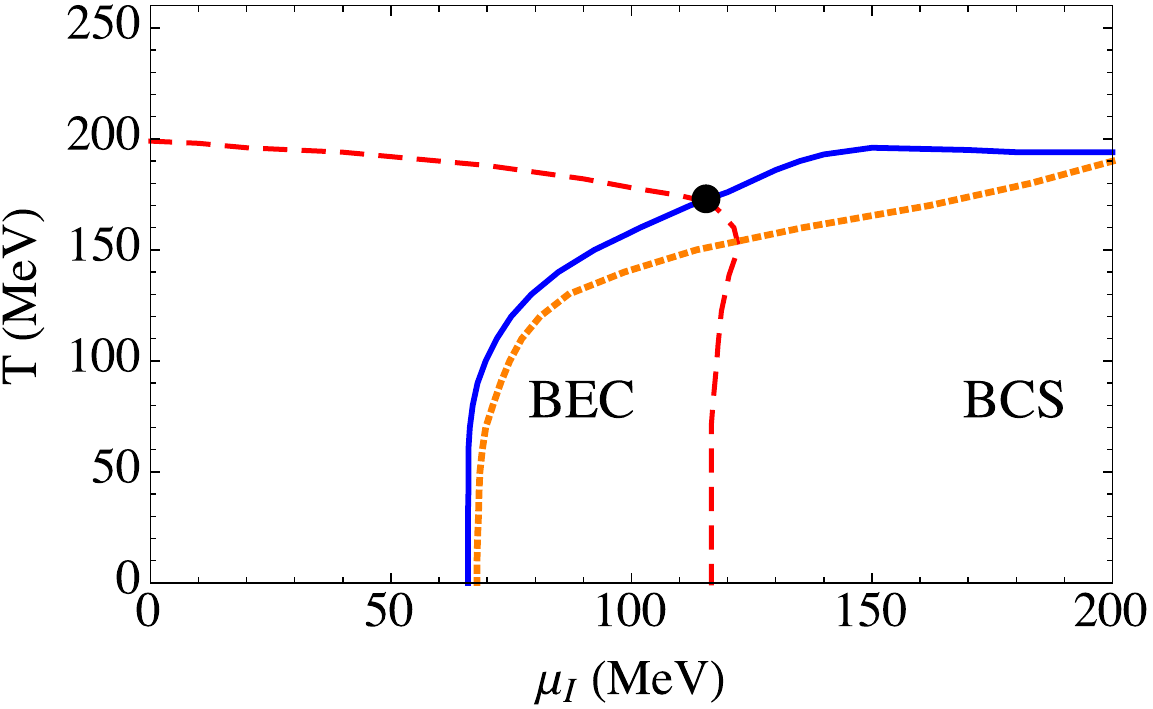}
\caption{The phase diagram of pion superfluidity in the $T$ and $\mu_I$ plane. The dotted and solid lines are respectively the phase boundaries in and beyond the potential approximation, and the dashed line indicates the BEC-BCS crossover in the pion superfluid and separates the quark-meson gas below and the quark gas above in the normal phase. }
\label{fig3}
\end{figure}

At potential level, we can calculate both the pion condensate in the superfluid phase and the meson masses in the normal phase, and then determine the phase transition line through either of the two conditions. The calculated boundary is shown as dotted line in Fig.\ref{fig3}, starting at $T=0$, $\mu_I=m_\pi(0,0)/2=67$ MeV. The critical temperature increases very fast in the beginning and then becomes smooth. The pion condensate is in the BEC type at low isospin density where the coupling between the pairing quark and anti-quark is strong and BCS type at high density where the coupling becomes weak~\cite{Sun:2007fc, He:2013gga}. The crossover line between the two types of condensates can be defined by the condition $m_{\Pi_+}(T,\mu_I)=2\left(m_q(T,\mu_I)-2\mu_I\right)=0$, namely $m_q(T,\mu_I)=\mu_I$ which indicates the opening of the decay channel $\Pi_+\to q\bar q$, see the dashed line in Fig.\ref{fig3}. Since $\Pi_+$ is the massless pion mode in the superfluid phase with $m_{\Pi_-}>m_{\pi_0}>m_{\Pi_+}=0$, there are no more bound states of quarks on the right-hand side of the crossover line. Above the phase transition line, the crossover line is continued by the constraint $m_{\pi_+}(T,\mu_I)=2\left(m_q(T,\mu_I)-\mu_I\right)$ which separates the meson-quark phase below and the quark phase above.

\begin{figure}[!hbt]\centering
\includegraphics[width=0.4\textwidth]{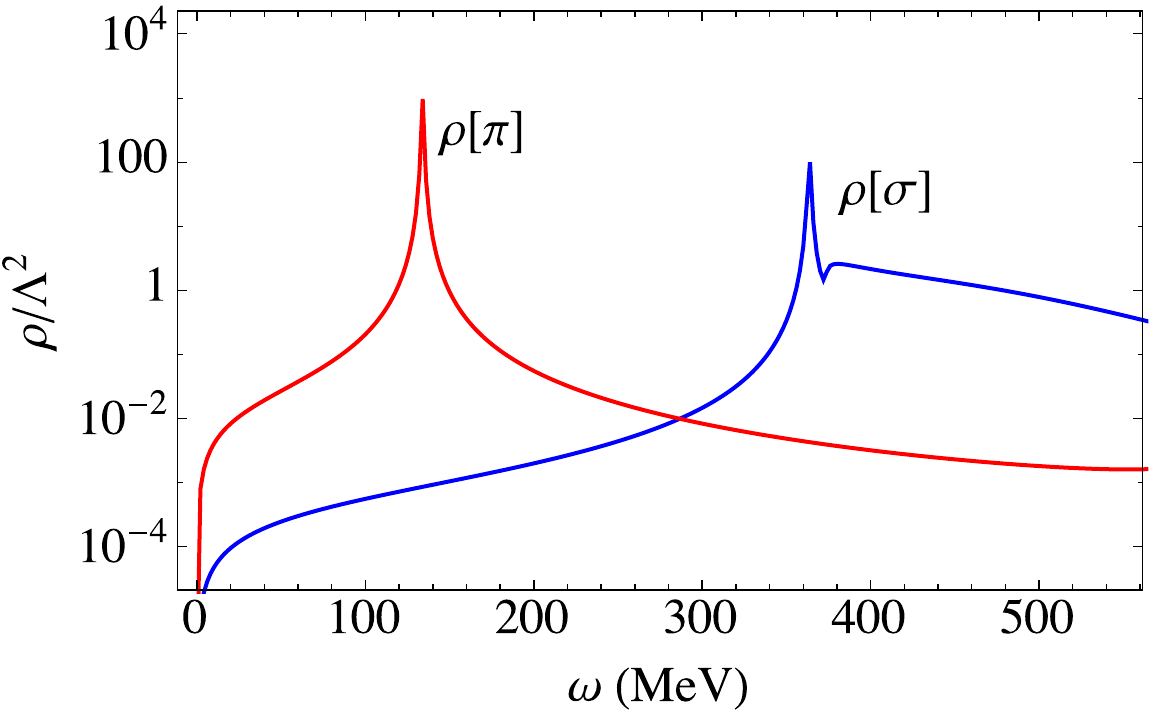}
\caption{The scaled $\pi$ and $\sigma$ spectral functions in vacuum. }
\label{fig4}
\end{figure}

Beyond the potential level, we study the meson two-point functions only in the normal phase, to avoid the complicated calculation due to the off-diagonal elements in the superfluid phase. In this case, we extract the mass of the soft mode $\pi_+$ from the location of the peak of the spectral function, $d\rho_{0,\omega}[\pi_+]/d\omega|_{\omega=m_{\pi_+}}=0$, and determine the phase boundary through the condition $m_{\pi_+}(T,\mu_I)=0$, see the solid line in Fig.\ref{fig4}. The two phase transition lines in and beyond potential approximation coincide at the starting point, but the thermal and quantum fluctuations included in the higher order vertex functions $\Gamma^{(3)}$ and $\Gamma^{(4)}$ enhance the critical temperature sizeably.

We now turn to the meson spectral functions in the normal phase. The spectral functions $\rho[\pi]$ and $\rho[\sigma]$ in vacuum in the limit of $k \to 0$ are shown in Fig.\ref{fig4}. Note that, we calculated the spectral functions in case B, the peak of $\rho[\pi]$ is located at $\omega=m_\pi$. If we calculate the pion spectral function in case A, the peak is located at $94$ MeV which is far from the well-known pion mass. The reason why we take case B is the following: when we go beyond the potential approximation and calculate the two-point functions, we should determine the initial condition at the same level.

To focus on the thermal excitation of the condensate and see the crossover from BEC to BCS pairing, we consider the $\pi_+$ spectral function $\rho_{k,\omega}[\pi_+]$ slightly above the phase transition line, where the isospin symmetry is restored and the Goldstone mode $\Pi_+$ becomes the soft mode $\pi_+$. From the change of the width with isospin chemical potential, we can see directly whether the soft mode $\pi_+$ is a tightly or loosely bound state. From Fig.\ref{fig1}, the two decay channels $\pi_+\to \sigma\pi_+$ and $\pi_+\to u\bar d$ control the continuous spectrum $\rho_{k,\omega}[\pi_+]$. The corresponding threshold energies are $\omega \ge m_\sigma+m_{\pi_+}$ for the meson channel and $\omega \ge 2m_q$ for the quark channel. Considering the fact that sigma meson is always heavy on the phase boundary with mass $m_\sigma > 300$ MeV and quarks are heavy in the chiral breaking phase but become light in the chiral restoration phase, the meson channel opens only at very high energy but the quark channel can open easily when the chemical potential is high enough ($m_q < 100$ MeV for $\mu_I > 130$ MeV).
\begin{figure}[!hbt]\centering
\includegraphics[width=0.4\textwidth]{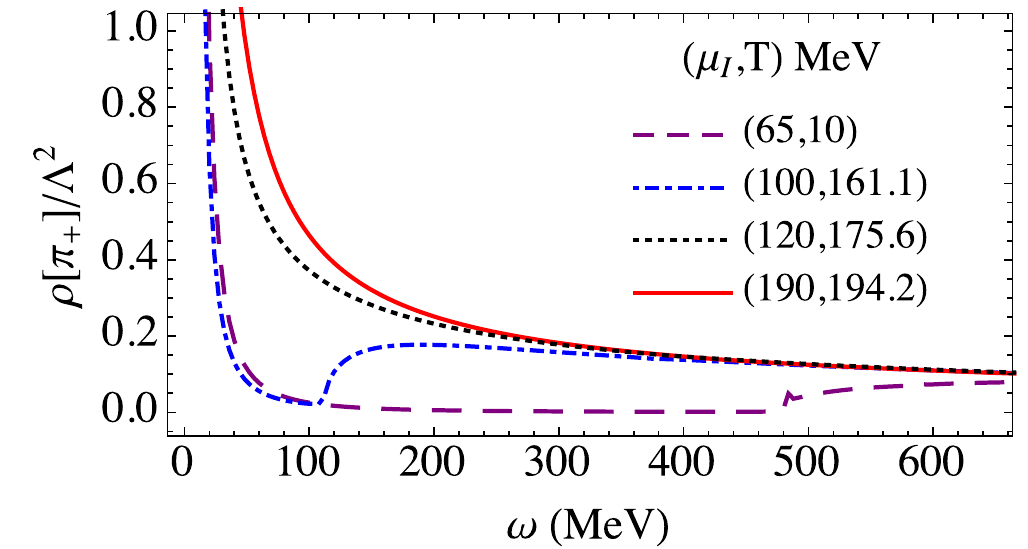}
\caption{The scaled spectral function $\rho_{k,\omega}[\pi_+]$ slightly above the superfluid boundary at different isospin chemical potential. }
\label{fig5}
\end{figure}

\begin{figure}[!hbt]\centering
\includegraphics[width=0.4\textwidth]{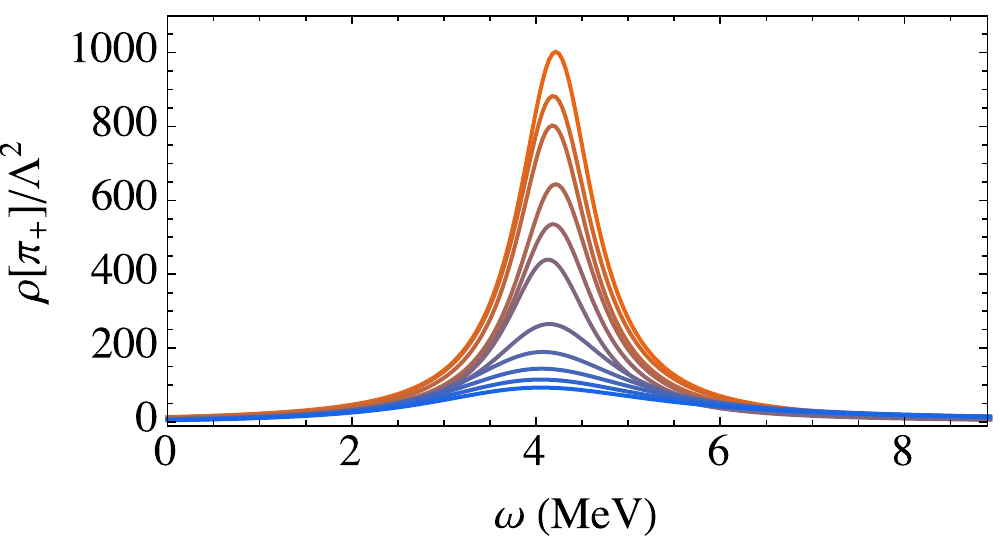}
\caption{The scaled spectral function $\rho_{k,\omega}[\pi_+]$ slightly above the superfluid boundary. The isospin chemical potentials are chosen to be $67,\ 70,\ 80,\ 90,\ 100,\ 110,\ 120,\ 130,\ 150,\ 170$ and $190$ MeV from top to bottom, and the corresponding temperatures are so chosen to have the same location of the peaks. }
\label{fig6}
\end{figure}

 Fig.\ref{fig5} shows the scaled $\pi_+$ spectrum at different $\mu_I$. Since the temperature is chosen to be a little bit above the pion superfluid boundary, the location of the peak, namely the $\pi_+$ mass, is very close to zero at any $\mu_I$. However, the continuous part is sensitive to the chemical potential. At low $\mu_I$ the meson channel leads to a bump at high energy, and at high $\mu_I$ the quark channel opens already at low energy. To more clearly see the BEC to BCS crossover, we choose more values of isospin chemical potential to calculate the spectral function. We fix all the peaks to be located at the same energy, see Fig.\ref{fig6}. From top to bottom, the width of the spectral function increases monotonously with the isospin chemical potential. This shows explicitly a change from tightly to loosely bound meson states, corresponding to the BEC to BCS crossover in the pion superfluid.

 \begin{figure}[!hbt]\centering
\includegraphics[width=0.4\textwidth]{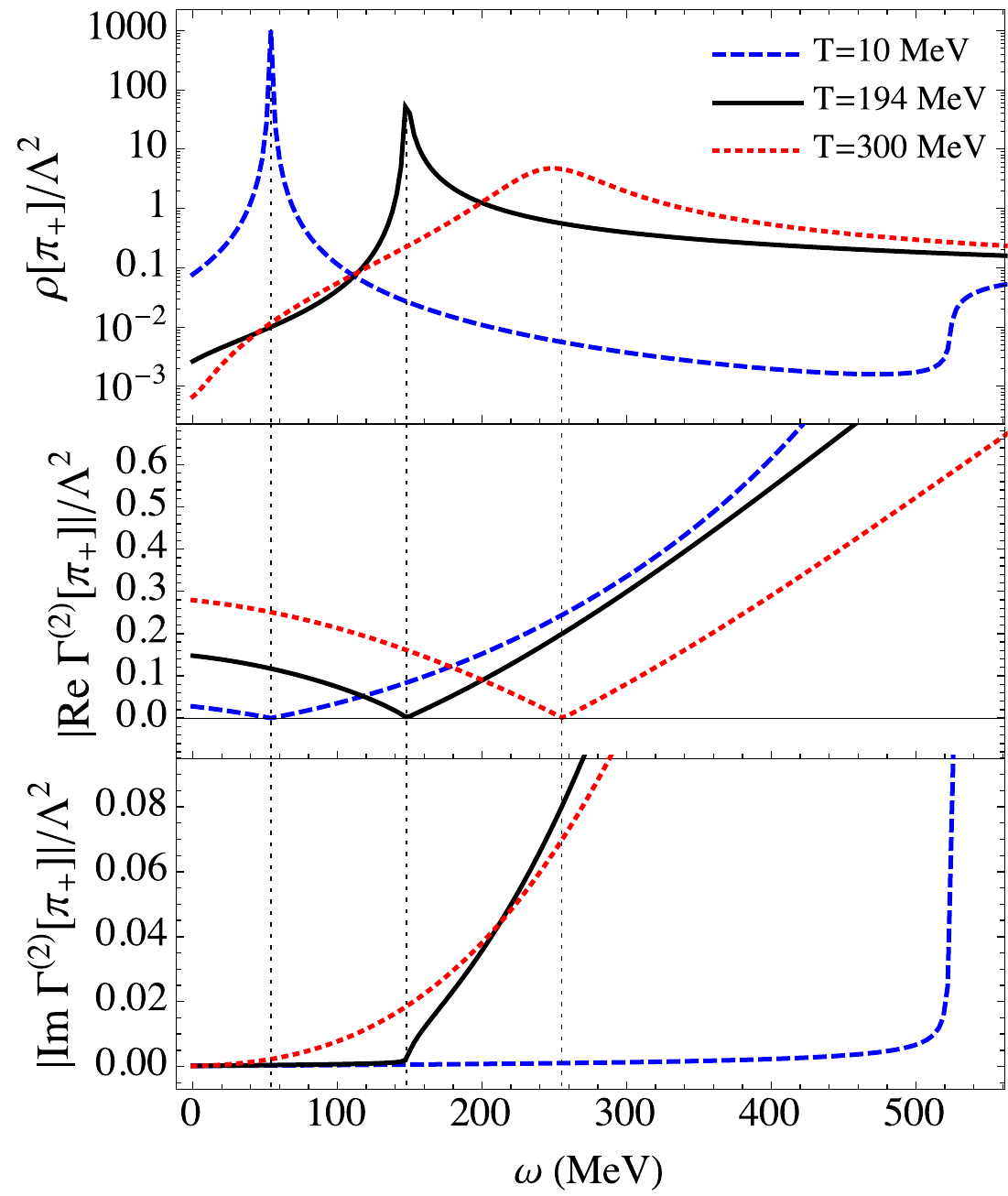}
\caption{The scaled spectral function (upper panel) and its real (middle panel) and imaginary (lower panel) parts for $\pi_+$ at fixed isospin chemical potential $\mu_I=40$ MeV and different temperature. }
\label{fig7}
\end{figure}
The transition from a mixed gas of mesons and quarks to a quark gas is expected to take place when the lightest meson mass is larger than two times the quark mass, which is denoted by the dashed line in the normal phase shown in Fig.\ref{fig3}. Beyond the potential level, we here show in Fig.\ref{fig7} the $\pi_+$ spectral function and its real and imaginary parts at fixed isospin chemical potential $\mu_I=40$ MeV and different temperature. At low temperature $T=10$ MeV which is much below the transition line in Fig.\ref{fig4}, mesons are tightly bound states of quarks with sharp peaks in spectral functions, see the dashed line in the upper panel of Fig.\ref{fig7}. The location of the peak is determined by the zero point of the real part, see the middle panel. Corresponding to the sharp peak, all the decay channels are closed and the imaginary part is zero, see the lower panel. The real lines in Fig.\ref{fig7} are the calculation at $T=194$ MeV which is exactly on the transition line. In this case, the decay channel $\pi_+\to q+\bar q$ starts to open, the peak in the spectral function is largely broaden and the imaginary part starts to have nonzero value. At $T=300$ MeV which is far above the transition line, the coupling between the quark and anti-quark is very weak, the peak becomes a bump and the imaginary part is already large.

\section{Summary}
\label{s4}

We investigated the phase diagram and the behavior of the Goldstone and soft modes of pion superfluid in the frame of a quark-meson model with functional renormalization group. Compared with solving only the flow equation for the effective potential which governs the thermodynamic property of the whole system, we focused on the flow equations for particle two-point functions which provide the information on particle propagation in the hot medium.

At potential level, we calculated the phase transition line of the pion superfluid in temperature and isospin chemical potential ($T-\mu_I$) plane. By considering the threshold condition for the meson decay channels controlled by the meson and quark curvature masses, we determined the crossover from BEC pairing at low $\mu_I$ to BCS pairing at high $\mu_I$ and the transition from quark-meson gas at low $T$ to quark gas at high $T$ in the normal phase. Beyond the potential level, we extracted the spectral function for the soft mode from its two-point function in the normal phase. By taking into account the Goldstone theorem, we redetermined the phase boundary of the pion superfluid and found a sizeable increase of the critical temperature, in comparison with the calculation at potential level. From the $\mu_I$ dependence of the spectral function slightly above the phase boundary, we clearly shown the change from tightly to loosely bound states of quarks, corresponding to the BEC-BCS crossover. Finally, from the $T$ dependence of the spectral function and its real and imaginary parts for the soft mode, we demonstrated again the transition of the system from a mixed quark-meson gas to a quark gas.

\noindent {\bf Acknowledgement:} We thank Dr. Ralf-Arno Tripolt for helpful discussions on numerical details. The work is supported by the NSFC and MOST grant Nos. 11335005, 11575093, 2013CB922000 and 2014CB845400 and Tsinghua University Initiative Scientific Research Program.

\begin{appendix}
\section{Loop functions}
\label{threshold}
With the boson and fermion occupation numbers and their derivatives,
\begin{eqnarray}
&& n_B(x)=\frac{1}{e^{x/T}-1},\quad n_F(x)=\frac{1}{e^{x/T}+1},\nonumber\\
&& n'_B(x)=\frac{dn_B(x)}{d x},\quad n'_F(x)=\frac{dn_F(x)}{d x},
\end{eqnarray}
the loop functions $J_{\phi}$ and $J_\psi$ in the flow equation for effective potential and $K_{\phi}$ with four-line vertices are explicitly expressed as
\begin{eqnarray}
J_\phi &=& \frac{k^4}{3\pi^2} \frac{1+2n_B(E_\phi-\mu_\phi)}{2E_\phi},\nonumber\\
J_\psi &=& \frac{k^4}{3\pi^2} \frac{1-n_F(E_\psi-\mu_\psi)-n_F(E_\psi+\mu_\psi)}{E_\psi},\nonumber\\
K_\phi &=& \frac{k^4}{3\pi^2}\frac{1}{4}\left[\frac{1+n_B(E_{\phi}-\mu_\phi)+n_B(E_{\phi}+\mu_\phi)}{E_{\phi}^3}-\frac{n'_B(E_{\phi}-\mu_\phi)+n'_B(E_{\phi}+\mu_\phi)}{E_{\phi}^2}\right].
\end{eqnarray}

After a straightforward but tedious calculation, the energy dependent loop functions $L_{\phi_i\phi_j}(p_0)$ for meson loops with three-line vertices can also be written as
\begin{widetext}
\begin{eqnarray}
L_{++}(p_0)
&=&\frac{k^4}{3\pi^2}\frac{1}{4}\bigg[\frac{12E_\alpha^2+p_0^2}{E_\alpha^3(4E_\alpha^2+p_0^2)^2}\left(1+n_B(E_\alpha+2\mu_I)+n_B(E_\alpha-2\mu_I)\right)\nonumber\\
&&\quad\qquad-\frac{1}{E_\alpha^2(2E_\alpha-i p_0)i p_0}n'_B(E_\alpha-2\mu_I)+\frac{1}{E_\alpha^2(2E_\alpha+i p_4)i p_0}n'_B(E_\alpha+2\mu_I)\bigg],\nonumber\\
L_{\sigma0}(p_0)
&=&\frac{k^4}{3\pi^2}\frac{1}{4}\bigg[\frac{E_\beta^2-(E_\alpha-ip_0)(3E_\alpha-ip_0)}{E_\alpha^3((E_\alpha-i p_0)^2-E_\beta^2)^2}(1+n_B(E_\alpha))+\frac{E_\beta^2-(E_\alpha+ip_0)(3E_\alpha+ip_0)}{E_\alpha^3((E_\alpha+i p_0)^2)^2-E_\beta^2}n_B(E_\alpha)\nonumber\\
&&\quad\qquad+\frac{1}{E_\alpha^2((E_\alpha-i p_0)^2-E_\beta^2)}n'_B(E_\alpha)+\frac{1}{E_\alpha^2((E_\alpha+i p_0)^2-E_\beta^2)}n'_B(E_\alpha)\nonumber\\
&&\quad\qquad+\frac{2}{E_\beta((E_\beta-ip_0)^2-E_\alpha^2)^2}n_B(E_\beta)+\frac{2}{E_\beta((E_\beta+ip_4)^2-E_\alpha^2)^2}(1+n_B(E_\beta))\bigg],\nonumber\\
L_{\sigma+}(p_0)
&=&\frac{k^4}{3\pi^2}\frac{1}{4}\bigg[\frac{E_\pi^2-(E_\sigma-ip_0-2\mu_I)(3E_\sigma-ip_4-2\mu_I)}{E_\sigma^3((E_\sigma-i p_0-2\mu_I)^2-E_\pi^2)^2}(1+n_B(E_\sigma))+\frac{n'_B(E_\sigma)}{E_\sigma^2((E_\sigma-i p_0-2\mu_I)^2-E_\pi^2)}\nonumber\\
&&\quad\qquad+\frac{E_\pi^2-(E_\sigma+ip_0+2\mu_I)(3E_\sigma+ip_4+2\mu_I)}{E_\sigma^3((E_\sigma+i p_0+2\mu_I)^2+E_\pi^2)^2}n_B(E_\sigma)+\frac{n'_B(E_\sigma)}{E_\sigma^2((E_\sigma+i p_0+2\mu_I)^2-E_\pi^2)}\nonumber\\
&&\quad\qquad+\frac{2n_B(E_\pi-2\mu_I)}{E_\pi((E_\pi-ip_0-2\mu_I)^2-E_\sigma^2)^2}+\frac{2(1+n_B(E_\pi+2\mu_I))}{E_\pi((E_\pi+ip_0+2\mu_I)^2-E_\sigma^2)^2}\bigg],\nonumber\\
L_{+\sigma}(p_0)
&=&\frac{k^4}{3\pi^2}\frac{1}{4}\bigg[\frac{E_\sigma^2-(E_\pi-ip_0+2\mu_I)(3E_\pi-ip_0+2\mu_I)}{E_\pi^3((E_\pi-i p_0+2\mu_I)^2-E_\sigma^2)^2}(1+n_B(E_\pi+2\mu_I))+\frac{n'_B(E_\pi+2\mu_I)}{E_\pi^2((E_\pi-i p_0+2\mu_I)^2-E_\sigma^2)}\nonumber\\
&&\quad\qquad+\frac{E_\sigma^2-(E_\pi+ip_0-2\mu_I)(3E_\pi+ip_0-2\mu_I)}{E_\pi^3((E_\pi+i p_0-2\mu_I)^2-E_\sigma^2)^2}n_B(E_\pi-2\mu_I)+\frac{n'_B(E_\pi-2\mu_I)}{E_\pi^2((E_\pi+i p_4-2\mu_I)^2-E_\sigma^2)}\nonumber\\
&&\quad\qquad+\frac{2n_B(E_\sigma)}{E_\sigma((E_\sigma-ip_4+2\mu_I)^2-E_\pi^2)^2}+\frac{2(1+n_B(E_\sigma))}{E_\sigma((E_\sigma+ip_4-2\mu_I)^2-E_\pi^2)^2}\bigg]
\end{eqnarray}
For fermion loops, divide $M^\sigma_{++}(p_4,\mu_I)$ into two terms for convenience, 
\begin{eqnarray}
M^\sigma_{++}(p_0)&=&M^0_{++}(p_0)+4m_\psi^2M^1_{++}(p_0),\nonumber\\
M^\pi_{++}(p_0)&=&M^0_{++}(p_0).
\end{eqnarray}
Loop functions $M^{\phi_i}_{\phi_j\phi_k}(p_0)$ related to fermion loops are given as following
\begin{eqnarray}
M^0_{++}(p_0)
&=&\frac{4k^4}{6\pi^2}\Bigg\{\frac{4E_q^2-p_4^2}{E_q(4E_q^2+p_4^2)^2}(n_F(E_q+\mu_I)+n_F(E_q-\mu_I)-1)-\frac{n'_F(E_q-\mu_I)}{2E_q(2E_q-ip_4)}-\frac{n'_F(E_q+\mu_I)}{2E_q(2E_q+ip_4)}\Bigg\}\nonumber\\
M^1_{++}(p_0)
&=&\frac{4k^4}{6\pi^2}\Bigg\{\frac{12E_q^2+p_4^2}{4E_q^3(4E_q^2+p_4^2)^2}(1-n_F(E_q+\mu_I)-n_F(E_q-\mu_I))+\frac{n'_F(E_q-\mu_I)}{4E_q^2(2E_q-ip_4)ip_4}-\frac{n'_F(E_q+\mu_I)}{4E_q^2(2E_q+ip_4)ip_4}\Bigg\}\nonumber\\
M^\pi_{+-}(p_0)
&=&\frac{4k^4}{6\pi^2}\Bigg\{\frac{1}{2E_q(2E_q+ip_4+2\mu_I)^2}(2n_F(E_q+\mu_I)-1)+\frac{1}{2E_q(2E_q-ip_4-2\mu_I)^2}(2n_F(E_q-\mu_I)-1)\nonumber\\
&&\quad\quad-\frac{1}{2E_q(2E_q+ip_4+2\mu_I)}n'_F(E_q+\mu_I)-\frac{1}{2E_q(2E_q-ip_4-2\mu_I)}n'_F(E_q-\mu_I)\Bigg\}
\end{eqnarray}
\end{widetext}
The other elements $L_{\phi_i\phi_j}(p_0)$ and $M^{\phi_i}_{\phi_j\phi_k}(p_0)$ appeared in Eq.(\ref{twopoint}) can be obtained from the symmetry among charged mesons,
\begin{eqnarray}
L_{--}(p_0)&=&L_{++}(p_0),\nonumber\\
L_{00}(p_0)&=&L_{++}(p_0)|_{\mu_I=0},\nonumber\\
L_{\sigma-}(p_0)&=&L_{\sigma+}(p_0)|_{\mu_I\to -\mu_I},\nonumber\\
L_{-\sigma}(p_0)&=&L_{+\sigma}(p_0)|_{\mu_I\to -\mu_I},\nonumber\\
M^\pi_{--}(p_0)&=&M^\pi_{++}(p_0),\nonumber\\
M^\pi_{-+}(p_0)&=&M^\pi_{+-}(p_0).
\end{eqnarray}
\end{appendix}

\bibliographystyle{unsrt}

\end{document}